\documentclass[epj-spec]{svjour}
\usepackage{graphics}
\begin{document}
\title{First direct measurement of jets in $\sqrt{s_{NN}}=200$ GeV Heavy Ion
  Collisions by STAR}

\author{Sevil Salur\inst{1}\fnmsep\thanks{\email{ssalur@lbl.gov}} (for the STAR Collaboration)}
\institute{Lawrence Berkeley National Laboratory, 1 Cyclotron Road MS-70R0319, Berkeley, CA 94720}
\abstract{We present the first measurement of reconstructed jets in ultra-relativistic heavy ion collisions. Utilizing the large coverage of the STAR Time Projection Chamber and Electromagnetic Calorimeter, we apply several modern jet reconstruction algorithms and background subtraction techniques and explore their systematic uncertainties in heavy ion events. The differential spectrum for inclusive jet production in central Au+Au collisions at $\sqrt {s_{NN} }= 200$ GeV is presented. In order to assess the jet reconstruction biases, this spectrum is compared with the jet cross section measured in $\sqrt{s}=200$ GeV p+p collisions scaled by the number of binary N-N collisions to account for nuclear geometric effects.
}

\maketitle

\section{Introduction}
\label{intro}

Strong suppression of inclusive hadron distributions and di-hadron correlations at high $p_{T}$ is observed at the Relativistic Heavy Ion Collider \cite{Shighpt,Phighpt}. This provides direct evidence for partonic energy loss in dense matter generated in ultra-relativistic heavy ion collisions. However, such measurements suffer well-known geometric biases due to the competition of energy loss and fragmentation - the leading particle spectrum is dominated by relatively low energy jets that happen to lose little energy in the medium and fragment into higher $p_{T}$ particles \cite{baier}. These biases can be removed if the underlying partonic kinematics are reconstructed in an unbiased way, independent of the fragmentation details - quenched or unquenched. An unbiased jet reconstruction measurement in heavy ion collisions would give access to the full spectrum of fragmentation topologies without geometric biases, enabling full exploration of quenching dynamics. In addition, fully reconstructed jets allow the measurement of qualitatively new observables such as jet shapes, fragmentation functions, and energy flow. 

In this article, we present the first systematic measurement of direct jet reconstruction in ultra relativistic heavy ion collisions.  We study the effects of high multiplicity underlying events of reconstructed jets by  applying several modern jet reconstruction algorithms and background subtraction techniques that are appropriate to heavy ion collisions. See \cite{jor} for the accompanying discussion about  jet fragmentation studies in heavy ion collisions also presented for the first time during the Hard Probes 2008 meeting. 

\section{Analysis and Techniques}
\subsection{Event Selection and Terminology}

This analysis utilizes Au+Au data at $\sqrt{s_{NN}} = 200$ GeV recorded by STAR for the minimum bias and enhnanced high $p_{T}$ collisions during the year 7 RHIC run. The most central (0-10\%) Au+Au collisions are selected, using the multiplicity measurement performed with the STAR Time Projection Chamber (TPC). Two event sets were analyzed, based on fast on-line trigger configurations: \\
(i) MB-Trig, (minimum-bias trigger) utilizing the coincidence between the two calorimeters at beam rapidity (Zero Degree Calorimeters) with signals in each greater than $\sim40\%$ of the most probable amplitude for a single neutron, and \\
(ii) HT-Trig (high-tower trigger) that satisfies the  MB-Trig conditions and the additional requirement of a 2-tower EMC clusters having at least a 7.5 GeV energy deposition. 

 
For this study we use 3 million MB-Trig events, corresponding to 300 thousand 0-10\% most central events. A total of 80 million MB-Trig events were recorded by STAR during the 2007 $Au+Au$ run, but at present only the event set we use have been fully reconstructed off-line. The HT-Trig is designed to enhance the recorded rate of high $p_{T}$ photons and electrons. It may also serve to enhance the recorded rate of jets, though potentially with a bias that we investigate in this study. The HT-Trig dataset, which samples the full integrated luminosity of Run 7, has been reconstructed in its entirety, corresponding to 500 $\rm \mu b^{-1}$.

In order to assess jet reconstruction energy resolution, background subtraction, efficiency and acceptance, we utilize Monte-Carlo model studies based on Pythia 8.107 \cite{pythia}. 
Pythia events with hight $E_{T}$ jets are generated in three different configurations:\\ 
(i) PyTrue:  Pythia isolated jets including all particles except neutrinos.  Jets are  reconstructed using the PYTHIA internal jet finder, PyCell, for the cone algorithm, and $FastJet$ for the sequential recombination algorithm. \\
(ii) PyDet:  Pythia isolated jets (parameterized detector response level) reconstructed using the algorithms we apply to real data. \\
(iii) PyEmbed:  PyDet that are embedded in a background of real Au+Au 0-10\% central events, with jets reconstructed using the algorithms we apply to real data. 

We use several jet reconstruction algorithms to study the systematics. See next section for the detailed discussion of the algorithms.  In all of these algorithms, only the highest energy jet per event is selected as the reconstructed jet.

\subsection{Jet Reconstruction Algorithms and Background Corrections}

Jets are reconstructed by combining charged particles from the STAR TPC and the neutral energy from the Electromagnetic Calorimeter (EMC).  TPC and EMC detectors cover full azimuth ($0 < \phi <2 \pi$)  and mid rapidity ($ -1< \eta <1 $) of the events.  Corrections are applied for double-counting of energy due to hadronic energy deposition in the EMC and to electrons.
Two kinds of jet reconstruction algorithms are utillized; seeded cone (leading order high seed cone (LOHSC)) and sequential recombination ($K_{T}$ and Cambridge/Aachen).  For an  overview of jet algorithms in high energy collisions, see \cite{blazey,salamtalk,catchment,jets,salurww} and references therein. In the following, we briefly discuss the two kinds of algorithms and the corresponding underlying event subtraction that we use. 

The main method of the cone algorithm is to combine particles in $\eta - \phi $ space with their neighbors within a cone of radius R ($R=\sqrt{ \Delta \phi ^{2}+ \Delta \eta^{2} }$).  To optimize the search and effectiveness of jet finding in $p+p$ and $e^{+}+e^{-}$ collisions, splitting, merging, and iteration steps are used. In order to avoid instabilities in cone-finding due to large heavy ion background, we use a simple seeded cone without iteration or split-merging steps, with cone radius $R=0.4$
and minimum seed of 4.6 GeV.  A relatively small cone size is chosen to suppress the underlying heavy ion background \cite{sarah,joren}.  In $p+p$ collisions $\sim 80$\% of the jet energy is observed to be within R$\sim0.3$ for 50 GeV jets in the Tevatron \cite{cdf}.  However, broadening of the jet fragmentation due to quenching in the medium formed in heavy ion collisions may reduce the fraction of the measured energy in a given cone size and needs further exploration. To  further reduce the heavy ion background, the minimum accepted transverse momentum of charged particles and the transverse energy of the calorimeter cells ($p_{T}^{cut}$) is  varied between 0.1 to 2 GeV.  This threshold cut does not remove all the background contamination on the jet energy and additional subtraction is needed. The residual background is corrected based on the out-of-cone energy for the same $p_{T}^{cut}$, averaged over the STAR acceptance but measured on an event-wise basis, and scaled to the cone area.

The sequential recombination algorithms  have been used extensively in the Tevatron as they are collinear and infrared safe \cite{cdf,kt}.  In these types of algorithms, arbitrary shaped jets are allowed to follow the energy flow resulting in less bias on the reconstructed jet shape than with cone algorithms \cite{catchment}.   The $FastJet$ package for sequential recombination algorithms was used for this analysis  \cite{fastjet,akn}. The $FastJet$ package includes $k_{T}$, Cambridge/Aachen, anti-$k_{T}$, and an interface to external jet finders such as SisCone  via a plugin mechanism.  For infrared and collinear safe algorithms  an active area ($A_{j}$) of each jet is estimated by filling an event with many very soft particles and then counting how many are clustered into a given jet.  If the underlying event is distributed uniformly in $\eta$ and $\phi$ then this noise density can be subtracted from the measured jet energy on an event-by-event basis to correct for the background energy underlying the jet. In simulations, this correction is observed to recover most of the jet's energy when they are reconstructed in pile up and heavy ion backgrounds \cite{catchment}.

\begin{figure}[h!]
\begin{center}
\resizebox{0.9\textwidth}{!}{%
	\includegraphics{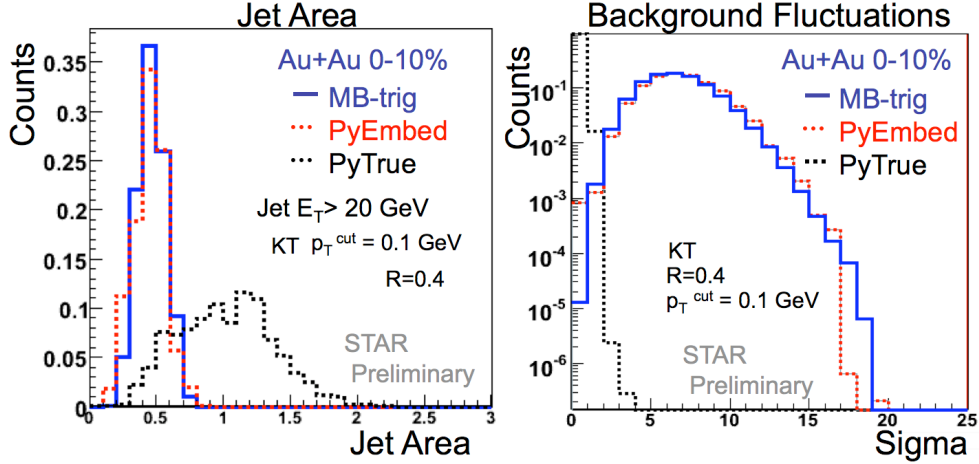} 
}	
\end{center}
\caption{Jet area and fluctuations are reconstructed utilizing the $FastJet$ algorithm \cite{catchment} for real jets in minimum bias triggered 0-10\% central $Au+Au$ collision (MB-trig),  in Pythia isolated jet events embedded in real central $Au+Au$ events (PyEmbed) and in Pythia isolated jet events (PyTrue).}
\label{fig:area}       
\end{figure}

Figure \ref{fig:area} shows the jet area and the fluctuations obtained directly from $FastJet$ framework with the $k_{T}$ algorithm for the PyTrue, PyEmbed and MB-Trig jets.  See the insets of the histograms for the applied parameters.  Pythia jets embedded in real Au+Au background events are observed to have the same area and fluctuations as jets from real Au+Au events.  This can be seen from the comparison of solid blue and dashed red histograms in Figure~\ref{fig:area}.  A reduction in the jet area and an increase in fluctuations are observed for the MB-Trig jets when compared with the PyTrue jets. This reduction of the jet area relative  to PyTrue is well understood for sequential recombination algorithms on theoretical grounds \cite{catchment}.

The inclusive jet spectrum from $p+p$ collisions at  $\sqrt s=200$ GeV ($p+p \;STAR$)  is used to make comparisons with the jet results in heavy ion collisions. Jets in $p+p$ collisions are measured the same way as in $Au+Au$ collisions, utilizing the STAR TPC and EMC and correcting for missing and double counted energy \cite{starpp}. However for the $p+p$ case, a mid-point  jet cone algorithm with splitting and merging steps is used.   The inclusive jet spectrum agrees well with the Next-to-leading order perturbative QCD calculation \cite{nlo}.

\subsection{Energy Resolution}

\begin{figure}[h!]
\begin{center}
\resizebox{0.96\columnwidth}{!}{
  \includegraphics{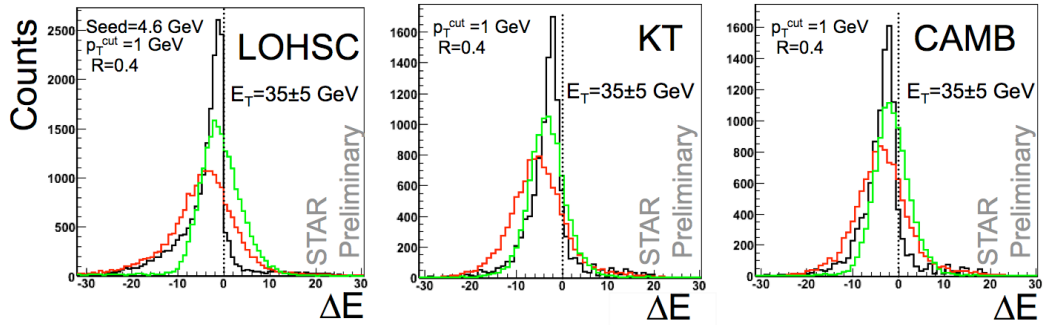} }
  \end{center}
\caption{Distributions showing energy resolution; black is  $\rm \Delta E =  E_{PyDet} - E_{PyTrue}$, red is $\rm \Delta E = E_{PyEmbed} - E_{PyTrue}$ 
and green is $\rm \Delta  E = E_{PyEmbed} - E_{PyDet}$.}
\label{fig:res}       
\end{figure}

The energy resolution for jet reconstruction with various algorithms has been studied with isolated jets simulated with Pythia \cite{pythia,yichun,mark}.  Figure \ref{fig:res} shows the event by event comparison 
of PyTrue, PyDet and PyEmbed from LOHSC, $ k_{T}$ and Cambridge/Aachen algorithms.  Applied cuts, jet energy and the algorithm used, are specified in the figures. For all three algorithms, a shift of median due to un-measured particles, primarily neutrons and $K^{0}_{L}$, and the applied  $p_{T}$ cut (hence loss of jet energy), is observed for the  
 $\rm \Delta E =  E_{PyDet} - E_{PyTrue} $ histograms.  Background  effects are simulated using  Pythia jets that are embedded in real $Au+Au$ events. The distributions in Figure~\ref{fig:res} is convoluted with the true jet spectrum to produce the observed jet spectrum.  
 The effect of the heavy ion background on the jet energy can be seen in the $\rm \Delta E=E_{PyEmbed}-E_{PyTrue}$ distributions. The algorithms show slightly different behavior: the LOHCS algorithm has more entries with $\rm \Delta E  >  0$  while the $k_{T}$ algorithm shows fewer entries at $\rm \Delta E  >  0$, but has average $\rm \Delta E  <  0$ , i.e. a overall oversubtraction of the background due to the back-reaction of the background (smaller average area, see Figure~\ref{fig:area} and \cite{catchment}).  A positive  $\rm \Delta E$ in this distribution can distort the measured inclusive jet spectrum substantially, increasing the apparent yield at high $E_{T}$ and resulting into a harder spectrum.  A correction to the spectrum must be applied to account for this effect.

\begin{figure}[h!]
\begin{center}
\resizebox{0.96\columnwidth}{!}{%
  \includegraphics{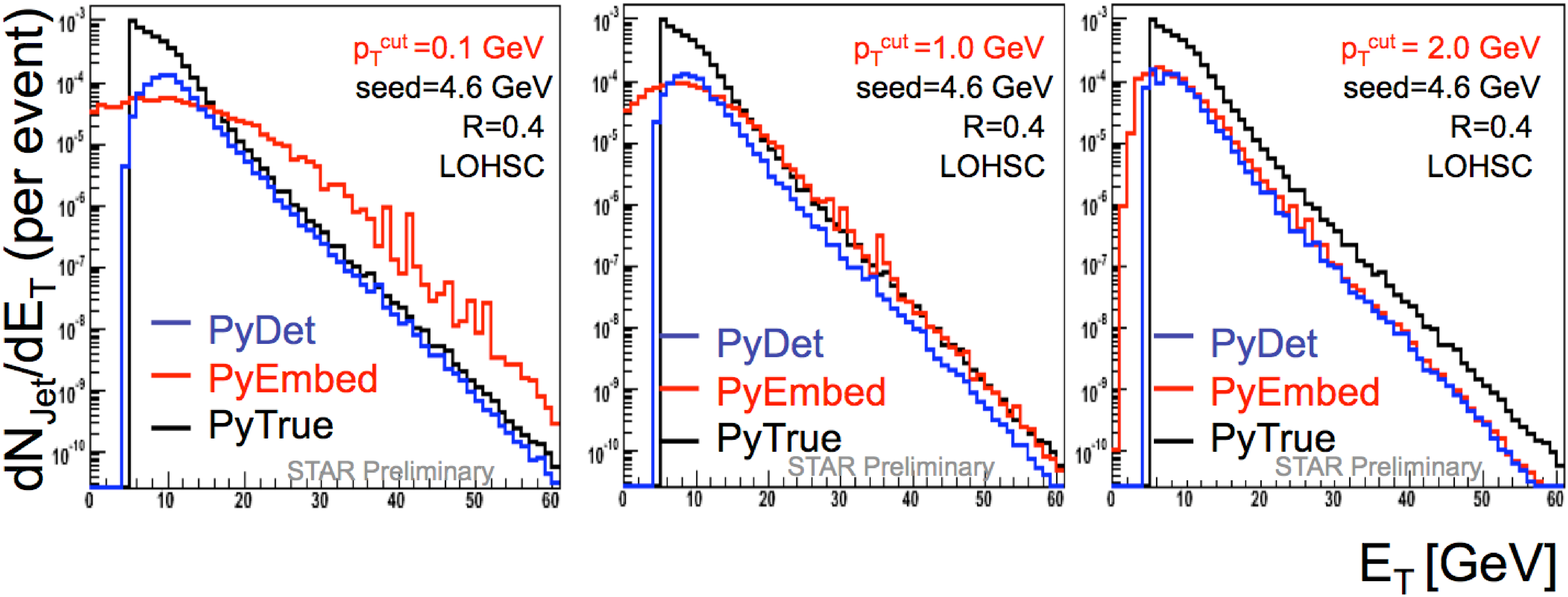} }
  \end{center}
\caption{Inclusive jet spectrum for PyDet, PyEmbed and PyTrue using the LOHSC algorithm with three different $p_{T}$ cuts on track momentum and calorimeter cell energy.   Note the lower threshold on generated jet energy $E_{T}^{PyTrue} > 5$ GeV, which affects the reconstructed spectrum up to $E_{T}=20$ GeV.}
\label{fig:reseff}       
\end{figure}

The influence of energy resolution on jet spectrum can be observed with PyDet, PyEmbed and PyTrue distributions shown in Figure \ref{fig:reseff} for the LOHSC algorithm for $p_{T}^{cut}$ of 0.1, 1 and 2 GeV.  In the smallest $p_{T}^{cut}$ case, a large difference between PyEmbed and PyDet is observed, while in the $p_{T}^{cut}=2$ GeV case when the  background contributions is the smallest, PyEmbed is similar to PyDet. The kick in the jet spectrum  for PyEmbed is observed in the $p_{T}^{cut}=0.1$ GeV,  as expected from the tail at positive $\Delta E$ due to large background in red distributions in Figure~\ref{fig:res}. When the $\rm p_{T}$ threshold is increased, the background fluctuations are reduced and the enhancement in the spectrum relative to the case without background is reduced to a negligible level. With this larger $p_{T}^{cut}=2$ GeV threshold requirement,  jet reconstruction in 0-10\% most central Au+Au collisions is similar to that of p+p collisions. However, while a reduction in the background fluctuations is achieved, a reduction in the measured jet energy ($\rm p_{T}^{cut}$ dependent bias) is introduced. Similar effects, though smaller in magnitude, are also observed for the $\rm K_{T}$ and Cambridge/Aachen algorithms.

\subsection{Jet Spectrum and Corrections}

\begin{figure}[h!]
\begin{center}
\resizebox{0.96\columnwidth}{!}{%
  \includegraphics{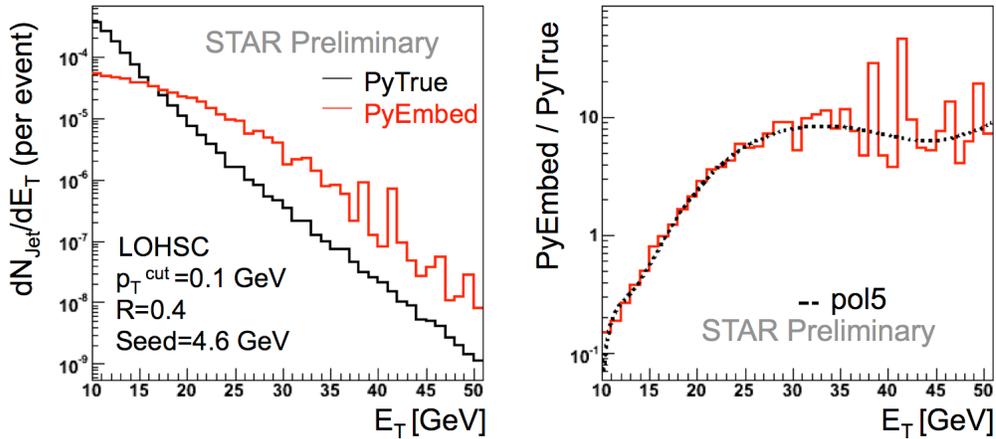} }
  \end{center}
\caption{The inclusive jet spectra from PyTrue and PyEmbed constructed with the LOHSC are shown on the left and their ratio is shown on the right.}
\label{fig:correction}       
\end{figure}

The correction factors of the jet spectrum are estimated using PyEmbed. The comparison of the spectrum of the jets of PyTrue and PyEmbed are shown in the left panel of Figure~\ref{fig:correction}.  The $\rm E_{T}$ dependent ratio of PyEmbed to PyTrue is shown in the right panel of Figure~\ref{fig:correction}.  A polynomial function fit to the distribution, shown as the dashed histogram, is used as a multiplicative correction to the inclusive spectrum. Table~\ref{tab:1} shows the inclusive jet spectrum correction factors for various $p_{T}^{cut}$ values.  For the sequential clustering algorithms, the correction factors are closer to unity.

\begin{table}
\caption{ 
Correction factors for the inclusive jet spectrum, for different reconstruction algorithms and values of $p_{T}^{cut}$. 
The range of values given indicates the correction factor variation from the lowest to highest jet $E_{T}$ shown in the figures.
}
\label{tab:1}       
\begin{center}
\begin{tabular}{l|l|l|l}
\hline\noalign{\smallskip}
$p_{T}^{cut}$ & LOHSC &  $k_{T}$  & CAMB \\
\noalign{\smallskip}\hline\noalign{\smallskip}
0.1 GeV & 0.2-10 & 1-4 & 2-6 \\
1 GeV & 0.2-1 & 0.7-1 & 1-2\\
2 GeV & 0.2-0.3 & 0.5-1 & 0.5-1\\
\noalign{\smallskip}\hline
\end{tabular}
\end{center}
\end{table}

The corrected inclusive jet spectrum for the LOHSC algorithm for the $p_{T}^{cut}=1$ GeV is presented in Figure~\ref{fig:spectra}.  The $p_{T}^{cut}=1$ GeV  is selected as it corresponds to correction factors close to unity. At this $p_{T}^{cut}$ the competing effects of energy loss due to momentum threshold cut and the kick up in the jet spectrum due to the positive tail of energy resolution cancel each other. The filled triangles are for the MB-Trig data set and are corrected for resolution, acceptance and efficiency.

\subsection{Comparison to $p+p$ Spectrum}

In order to assess the biases in the jet spectrum reconstructed in central Au+Au collisions, we compare to the spectrum measured in p+p collisions. To account for nuclear geometric effects we scale the p+p spectrum by $\rm N_{Binary}$, the number of binary nucleon+nucleon collisions equivalent to a central Au+Au collisions, as calculated by a Glauber model \cite{glauber}. The cross section for hard processes is expected to scale with $\rm N_{Binary}$ if no nuclear effects are present. In the case of jet reconstruction, $\rm N_{Binary}$ scaling is expected if the reconstruction is unbiased, i.e. the jet energy is recovered independent of the particular mode of fragmentation, even in the presence of strong jet quenching. The $N_{Binary}$ scaled  jet spectrum from $p+p$ collisions is shown in filled squares \cite{starpp}.   The yellow band represents the systematic uncertainity of the $p+p$ jet measurement.  Heavy ion jet spectrum is observed to agree with $N_{Binary}$ scaled p+p measurement within the $\sim 50 \%$ systematic uncertainty of the normalization.

\begin{figure}[h!]
\begin{center}
\resizebox{0.5\textwidth}{!}{%
	\includegraphics{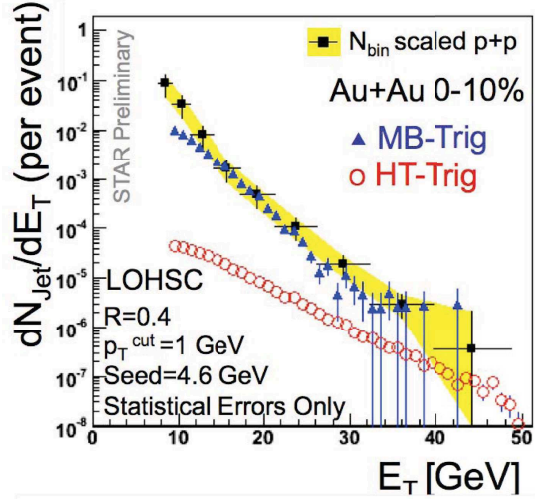} 
}	
\end{center}
\caption{Jet yield per event vs $E_{T}$ for 0-10\% central $Au+Au$ collisions, compared to the distribution from $p+p$ collisions scaled by  $\rm N_{Binary}$ \cite{starpp}.   
Triangle symbols are from MB-Trig and corrected for efficiency, acceptance and energy resolution. Open circles are from HT-Trig and not corrected for trigger bias. Only statistical error bars are shown for the $Au+Au$ data. Filled black squares are the distribution from $p+p$ collisions, scaled by $N_{Binary}$.  Band represents the systematic uncertainty of the $p+p$ measurement. }
\label{fig:spectra}       
\end{figure}

Open circles  in Figure~\ref{fig:spectra} show uncorrected jet spectrum from HT-Trig data which is substantially lower then the corrected MB-Trig spectrum.  The correction factors of energy resolution, efficiency and acceptance for the HT-Trig are  expected to be small.  A large trigger bias due to the additional 7.5 GeV energy deposition in EMCAL requirement fot HT-Trig relative to MB-Trig is seen to persist to at least to 30 GeV.  Further statistics of MB-Trig data is needed to assess the bias at high $p_{T}$. 

\begin{figure}[h!]
\resizebox{0.99\textwidth}{!}{%
	\includegraphics{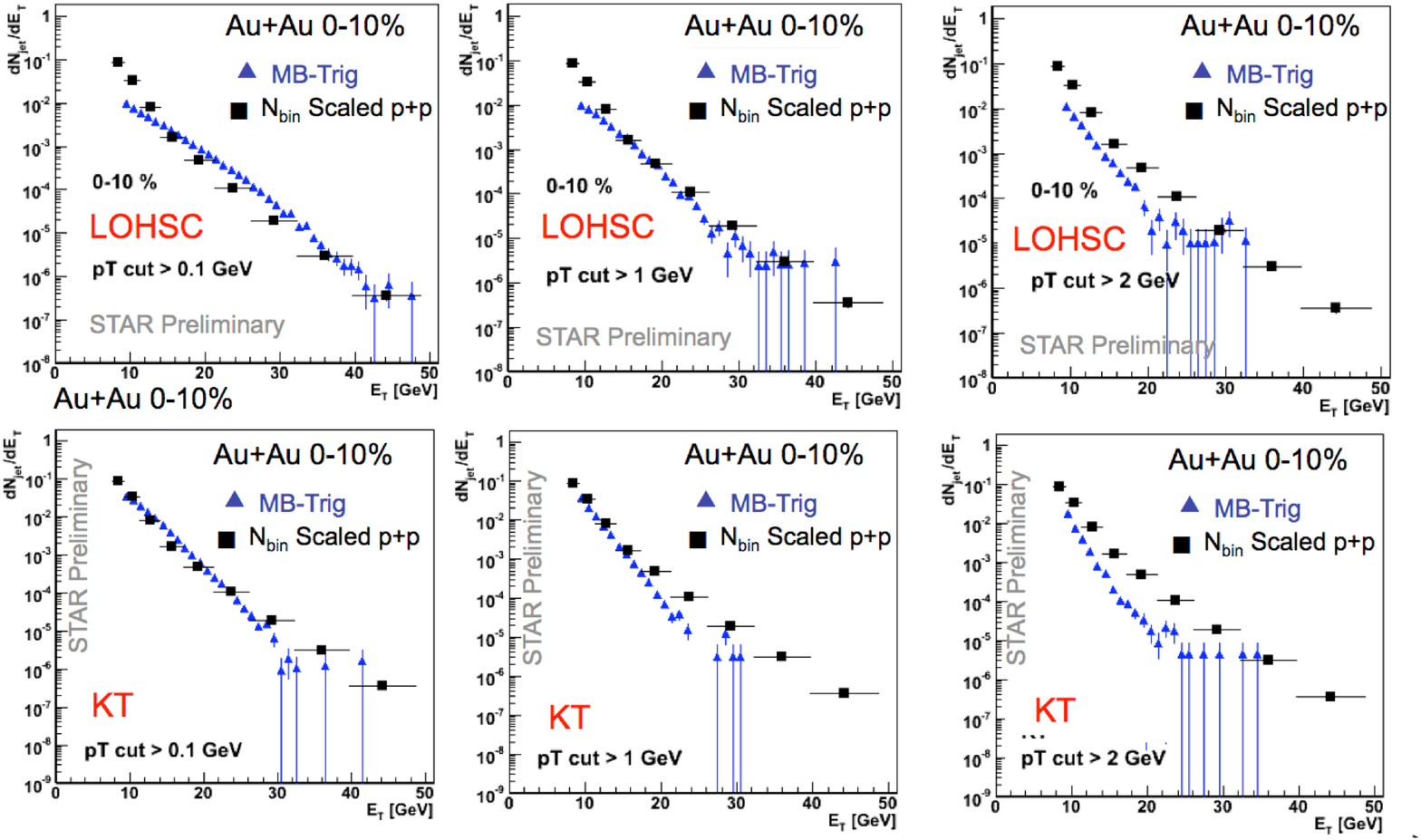} 
}	
\caption{The $p_{T}$ threshold dependent comparison of inclusive jet spectra in heavy ion and p+p collisions.  The filled triangle symbols are from MB-Trig and filled squares are from $N_{Binary}$ scaled $p+p$ collisions. Upper panel is for the jet yield per event vs $E_{T}$ reconstructed with LOHSC algorithm and the lower panel is for the same distributions with the $k_{T}$ algorithm.  In addition to $p_{T}$ cuts shown in the figures, seed is 4.6 GeV for the LOHSC and R is set to 0.4 for both algorithms. Only statistical error bars are shown. }
\label{fig:panelall}       
\end{figure}

Figure~\ref{fig:panelall} shows the comparison of inclusive jet spectra from the MB-Trig $Au+Au$ data and the $N_{Binary}$ scaled p+p for the variations of $p_{T}$ threshold cut for LOHSC and $k_{T}$ algorithms. 
As can be seen from this Figure, the agreement between $N_{Binary}$ scaled p+p and MB-Trig measurement is good for the lowest value of $p_{T}$ cut. It is also seen to be  poorer with  the larger $p_{T}$ threshold cut. This suggests that $p_{T}^{cut}$ introduces biases which are not fully corrected using Pythia as the fragmentation model, and may be an indirect indication of modified fragmentation due to quenching. 

\section{Summary and Outlook}

We have presented the first full reconstruction of jets in 0-10\% most central heavy ion collisions at RHIC energies.   Systematics of the underlying heavy ion background subtraction is studied via utilizing various algorithms and consideration of the jet area. The $\rm N_{Binary}$ scaling is observed for the least-biased cuts with the given $\sim 50\%$ systematic uncertainty of the $p+p$ jet spectrum measurement.   
An unbiased jet reconstruction in heavy ion collisions with MB-Trig data appears to be feasible. However, spectrum corrections are currently based on model calculations using Pythia fragmentation. This aspect, together with background subtraction techniques, spectrum variations due to cuts and reconstruction algorithms, must be investigated further in order to assess the systematic uncertainties of this measurement.

The LHC will begin operations in 2008, with the first heavy ion run expected in late 2009. At the LHC the heavy ion background is expected to be larger than at RHIC, but there will be copious production of very energetic jets, well above background \cite{peter}. The large kinematic reach at the LHC may provide sufficient lever-arm to map out the QCD evolution  of jet quenching \cite{solan}.  The comparison of  full jet measurements in the different physical systems generated at
RHIC and the LHC will provide unique and crucial insights into our understanding of jet quenching and the nature of hot QCD matter.


\end{document}